\documentclass[reprint, twocolumn,superscriptaddress, amsmath,amssymb, aps, pra]{revtex4-1}
\usepackage{ifpdf}
\pdfoutput=1

\usepackage[T2A]{fontenc}
\usepackage[english]{babel}
\usepackage{amsmath}
\usepackage{pdfsync}
\usepackage{graphicx}
\usepackage{hyperref}

\newcommand{\bs}{\begin{split}}
\newcommand{\es}{\end{split}}

\newcommand{\ket}[1]{\vert{#1}\rangle}
\newcommand{\vect}[1]{\boldsymbol{#1}}
\newcommand{\mcm}{$\mu$m}

\begin{document}


\title{Broadband biphotons in a single spatial mode}


\author{K.\, G.\, Katamadze}
\email[]{k.g.katamadze@gmail.com}
\affiliation{M.\, V.\, Lomonosov Moscow State University, 119992, Moscow, Russia}
\affiliation{Institute of Physics and Technology, Russian Academy of Sciences, 117218, Moscow, Russia}

\author{N.\, A.\, Borshchevskaya}
\affiliation{M.\, V.\, Lomonosov Moscow State University, 119992, Moscow, Russia}

\author{I.\, V.\, Dyakonov}
\affiliation{M.\, V.\, Lomonosov Moscow State University, 119992, Moscow, Russia}

\author{A.\, V.\, Paterova}
\affiliation{M.\, V.\, Lomonosov Moscow State University, 119992, Moscow, Russia}

\author{S.\, P.\, Kulik}
\affiliation{M.\, V.\, Lomonosov Moscow State University, 119992, Moscow, Russia}


\date{\today}

\begin{abstract}
We demonastrate the experimental technique for generating a spatial single-mode broadband biphoton field.
The method is based on a dispersive optical element which precisely tailors the structure of the type-I SPDC frequency angular spectrum in order to shift different spectral components to a single angular mode.
Spatial mode filtering is realized by coupling biphotons into a single-mode optical fiber.
\end{abstract}

\pacs{}
\keywords{Biphotons, SPDC, broadband, spectrum}

\maketitle

\section{Introduction}
Progress in preparation and manipulation of quantum states of light is boosted by development of quantum communication and quantum computation. Spontaneous parametric down-conversion (SPDC) \cite{Belinsky_Klyshko_1994} is the most efficient and widespread source of correlated photon pairs (biphotons).
Phenomenologically, one may describe the SPDC process as a spontaneous decay of pump photon \textit{p} traveling through a $\chi^{(2)}$ nonlinear medium into a pair of photons conventionally referred to as the signal \textit{s} and idler \textit{i}. Emerged photon pairs obey the energy conservation law and phase matching relation:

\begin{align}\label{eqn::phase_matching}
&\omega_{p} = \omega_{s} + \omega_{i},\\
&\vec{k}_p = \vec{k}_s + \vec{k}_i + \vec{\Delta}.
\end{align}

\begin{figure*}[t]
	\centering
		\includegraphics{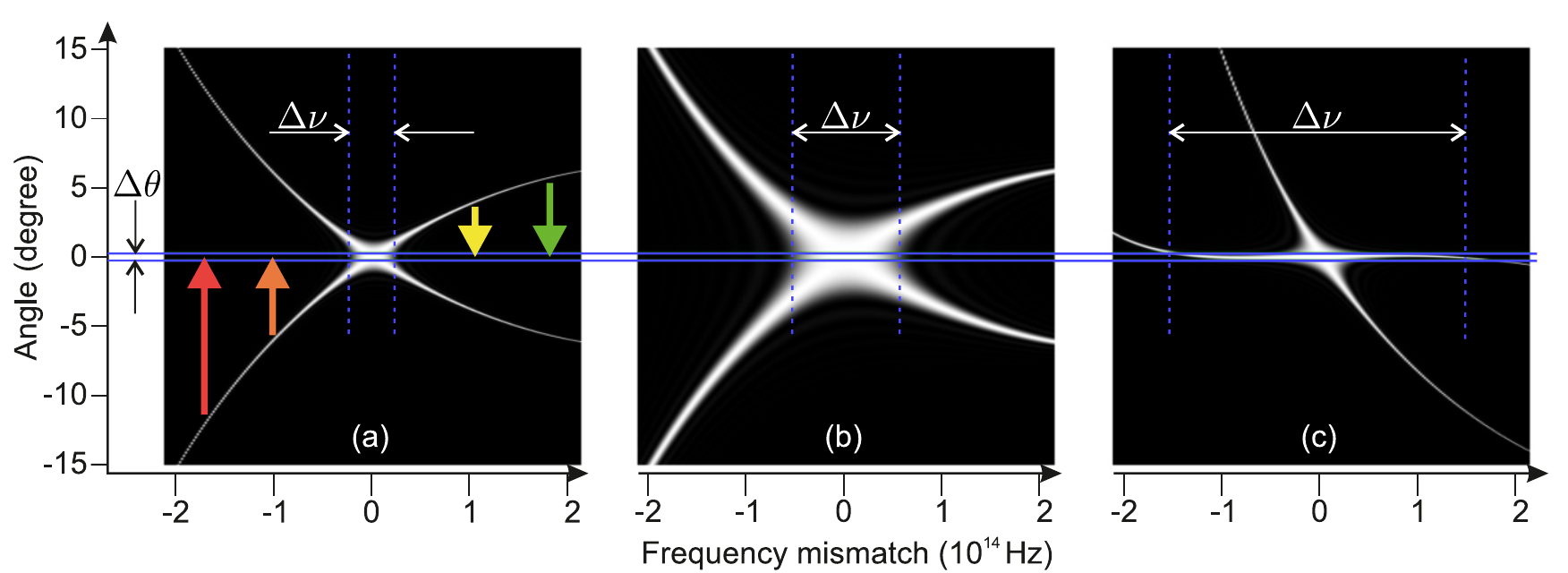}
	\caption{(Color online) The type-I collinear degenerate frequency-angular SPDC spectra, calculated for the 2~mm BBO crystal (zero on the frequency scale corresponds to the central frequency $\nu_o=\omega_p/4\pi$). (a)~The initial spectrum. (b)~The typical view of the spectrum for usual broadening techniques. (c)~The spectrum, transformed by an angular dispersion. The frequency bandwidth $\Delta \nu$ corresponds to the collection angular range $\Delta \theta$.}
	\label{fig:XXX}
\end{figure*}

Here, $\omega_{p,s,i}$ refer to the pump, signal and idler frequencies respectively, $\vec{k}_{p,s,i}$ -- corresponding wave vectors, and $\vec{\Delta}$ is a phase mismatch responsible for  the decay probability \cite{Belinsky_Klyshko_1994}. In the present work we will concentrate on investigating the frequency and angular characteristics of biphotons generated in the SPDC process. A biphoton quantum state may be expressed as

\begin{align}\label{eqn::biphoton_quantum_state}
\ket{\Psi} = F\left(\omega_{s},\theta_{s},\omega_{i},\theta_{i}\right)a^{\dagger}_{s}\left(\omega_{s},\theta_{s}\right)a^{\dagger}_{i}\left(\omega_{i},\theta_{i}\right)\ket{vac},
\end{align}
where $a^{\dagger}_{s}\left(\omega_{s},\theta_{s}\right)$ and $a^{\dagger}_{i}\left(\omega_{i},\theta_{i}\right)$ represent photon creation operators in frequency modes $\omega_{s,i}$ and angular modes $\theta_{s,i}$,
 and $F\left(\omega_{s},\theta_{s},\omega_{i},\theta_{i}\right)$ is a spectral amplitude of the biphoton. Relation between angular and frequency modes of a pair in monochromatic plane-wave pump approximation is given by:

\begin{align}
n(\omega_{i})\omega_{i}\sin{\theta_{i}} = n(\omega_{s})\omega_{s}\sin{\theta_{s}},
\end{align}
where $n(\omega)$ is refractive index.
Hence, the two-photon spectral amplitude $F\left(\omega_{s},\theta_{s},\omega_{i},\theta_{i}\right)$ reduces to a function of two variables $\tilde{F}\left(\omega_{s},\theta_{s}\right)$.
The typical X-shape of the frequency-angular spectrum of the biphoton field $|\tilde{F}\left(\omega_{s},\theta_{s}\right)|^2$ 
generated under collinear degenerate type-I phase-matching conditions is depicted in Fig.\ref{fig:XXX}~à.
For applications purposes a spatial single-mode biphoton field generation regime is preferential. A pump beam waist, which is typically
$ W \gtrsim 50$~$\rm{\mu m}$,  defines photon pair emission cone $\Delta\theta=\frac{2\lambda}{\pi W}$. This relation implies a narrow angular emission range in the single-mode generation regime.

Spectral broadening of the single spatial mode biphoton field remains a really significant puzzle among a wide list of biphoton spectrum control problems. The broadband biphotons exhibit ultra-short temporal correlations \cite{Belinsky_Klyshko_1994, Caspani_Brambilla_Gatti_2010, Chekhova_2002}. Their correlation time $\Delta \tau \sim 1/\Delta\nu$.
Ultra-correlated biphotons can be utilized in metrological applications such as distant clock synchronization \cite{Valencia_Scarcelli_Shih_2004}, quantum optical coherence tomography (QOCT) \cite{Carrasco_Torres_Torner_Sergienko_Saleh_Teich_2004a, Nasr_Saleh_Sergienko_Teich_2003} and quantum interferometric optical lithography \cite{Boto_Kok_Abrams_Braunstein_Williams_Dowling_2000, DAngelo_Chekhova_Shih_2001}. 

Besides, spectral broadening increases the degree of biphoton  entanglement \cite{Brida_Caricato_Fedorov_Genovese_Gramegna_Kulik_2009}, which can be exploited in quantum communication tasks. A quantum information can be directly encoded in frequency bins \cite{Bessire_Bernhard_Feurer_Stefanov_2014, Olislager_Woodhead_Phan_Huy_Merolla_Emplit_Massar_2014} or the broad spectrum can be used for a wavelength-multiplexed entanglement distribution \cite{Lim2008a, Lim2008, Chapuran2009, Herbauts2013, Donohue2014}. 

Also biphoton spectrum broadening increases a maximum pair generation rate $R_{max}\sim \Delta \nu$ \cite{Peer2005} enabling an extremely high-speed quantum communication.


Typically a biphoton spectral bandwidth is limited by an interaction medium dispersion ${\Delta}(\omega)<2\pi/L$, where $L$ is an interaction length. Hence, additional biphoton spectral broadening methods should be developed.
Homogeneous spectral broadening can be achieved by using the special nonlinear media with zero dispersion \cite{Peer2006, ODonnell2007, Shaked2014, Bisht_Shimizu_2015} ($\Delta \tau=5$~fs \cite{ODonnell2007} and $\Delta\nu=100$~THz \cite{Shaked2014} were obtained for type-I SPDC). 
Reducing nonlinear interaction length, one can generate 100~THz bandwidth type-II SPDC \cite{Dauler_Jaeger_Muller_Migdall_Sergienko_1999} and 150~THz type-I \cite{Katamadze_Borshchevskaya_Dyakonov_Paterova_Kulik_2013}, but this method essentially decreases an intensity of biphoton radiation.
Pump beam focusing \cite{Carrasco_Nasr_Sergienko_Saleh_Teich_Torres_Torner_2006, Carrasco_Sergienko_Saleh_Teich_Torres_Torner_2006, Carrasco_Torres_Torner_Sergienko_Saleh_Teich_2004b} and 
pump spectrum broadening \cite{Baek_Kim_2009, Nasr_Giuseppe_Saleh_Sergienko_Teich_2005} also enable the broadband biphoton generation with a bandwidth up to 83~THz \cite{Nasr_Giuseppe_Saleh_Sergienko_Teich_2005}.
One more method of homogeneous SPDC broadening is based on the modifying a crystal's effective dispersion by means of a pump pulse-front tilt combined with a spatial walk-off \cite{Hendrych_Shi_Valencia_Torres_2009}. A 44~THz bandwidth type-II SPDC spectrum was measured and a 197~THz bandwidth ($\Delta\tau=6.4$~fs) spectrum was theoretically predicted for type-I SPDC.

Biphoton generation in inhomogeneous nonlinear media (a set of crystals \cite{Okano_Okamoto_Tanaka_Subashchandran_Takeuchi_2012}, chirped crystals \cite{Carrasco_Torres_Torner_Sergienko_Saleh_Teich_2004a, Nasr_Carrasco_Saleh_Sergienko_Teich_Torres_Torner_Hum_Fejer_2008}, crystals with a spatially modulated refractive index \cite{Kalashnikov_Katamadze_Kulik_2009, Katamadze_Paterova_Yakimova_Balygin_Kulik_2011, Katamadze_Kulik_2011}) leads to inhomogeneous spectral broadening.
Up to 154~THz bandwidth of the SPDC spectrum was obtained \cite{Katamadze_Kulik_2011}, but such inhomogeneous broadening inevitably leads to a Fourier-unlimited field. 

All listed methods aim to increase parameter range (an area on the $\omega$-$\theta$ coordinate plane) satisfying the phase-matching conditions which inevitably leads to a proportional drop in the spectral intensity. The typical view of the frequency-angular spectrum for broadened (namely by reducing a crystal length \cite{Katamadze_Borshchevskaya_Dyakonov_Paterova_Kulik_2013}) type-I SPDC is shown in Fig.\ref{fig:XXX}~b.

It was demonstrated in \cite{Caspani_Brambilla_Gatti_2010, Gatti_Brambilla_Caspani_Jedrkiewicz_Lugiato_2009} that the up-conversion process evinces an ultra-short correlation time ($\Delta\tau=4.4$~fs)
 provided that biphotons were emitted into a wide angular range and then refocused onto the second nonlinear medium. However, collecting a wide angular range $\Delta \theta$ results in essentially a multimode biphoton generation regime, which is highly undesirable for long-distance (fiber or free-space) optical communication, microscopy and interferometric applications. Hence, the spatial single-mode broadband biphoton generation technique is a subject of endeavor for high-dimensional quantum-state engineering problems.

We have to note that the spatial single-mode biphoton generation problem (leaving spectral properties aside) has been thoroughly studied (for example \cite{Bovino_Varisco_MariaColla_Castagnoli_DiGiuseppe_Sergienko_2003, Kurtsiefer_Oberparleiter_Weinfurter_2001, Ling_Lamas-Linares_Kurtsiefer_2008, Ljunggren_Tengner_2005}). Generally, the solution is reduced to optimization of biphoton coupling conditions to a single-mode fiber. This implies maximization of single photon and coincidence counting rates and their ratio (the single to coincidence counting rate ratio is a key parameter in terms of pure single-photon state preparation). Our efforts were solely concentrated on achieving the maximal spectral width of a spatial single-mode biphoton source.

\section{Method description}


In spite of the fact that the total biphoton spectrum is broad both in frequency and angle, there is one-to-one correspondence between angles and frequencies due to phasematching.
We propose to reconfigure the SPDC spectrum by means of an angular dispersive optical element, shifting different spectral components to a single angular mode (Fig.~\ref{fig:XXX}~c). This method doesn't decrease the spectral intensity of SPDC radiation and significantly increases the integral intensity in the target angular mode.

\begin{figure}
	\centering
		\includegraphics{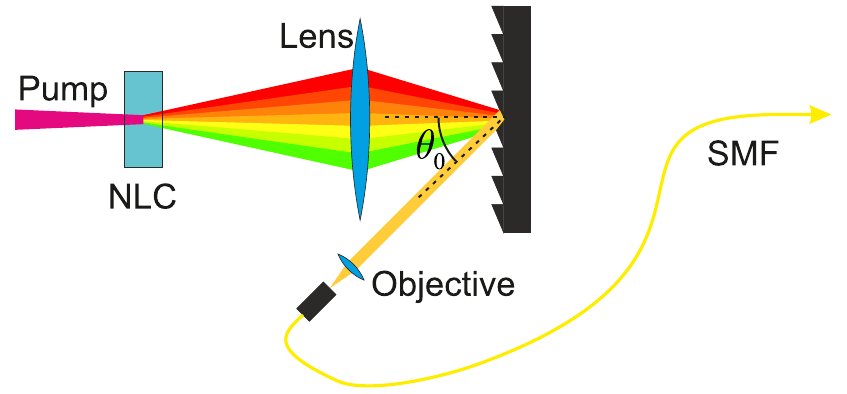}
	\caption{(Color online) The principal scheme of the experimental setup}
	\label{fig:Principal_setup}
\end{figure}

The principal scheme of the experimental setup is depicted in Fig.~\ref{fig:Principal_setup}. Different biphoton spectral components propagate in different angular modes. The lens system focuses emission onto an angular dispersive element (diffraction grating). Such a system transforms the SPDC angular spectrum to provide congruence with the grating dispersion curve. After diffracting from the grating, the majority of spectral components correspond to the same angular modes and the radiation is coupled by the objective to a single-mode fiber.

We have to point out that an analogous method was previously studied for optical parametric amplifier bandwidth broadening (for example \cite{Cardoso_Figueira_2004, Cardoso_Figueira_2005, Piskarskas_Stabinis_Pyragaite_2010, Stabinis_Krupic_2007, Yamane_Tanigawa_Sekikawa_Yamashita_2008}). However, it has never been utilized in biphoton generation applications. Nevertheless, biphoton spectral broadening using an angular dispersion module was demonstrated in \cite{Hendrych_Shi_Valencia_Torres_2009}, where the pump pulse-front tilt was implemented by means of a diffraction grating.

\section{Model}

\begin{figure}
	\centering
		\includegraphics{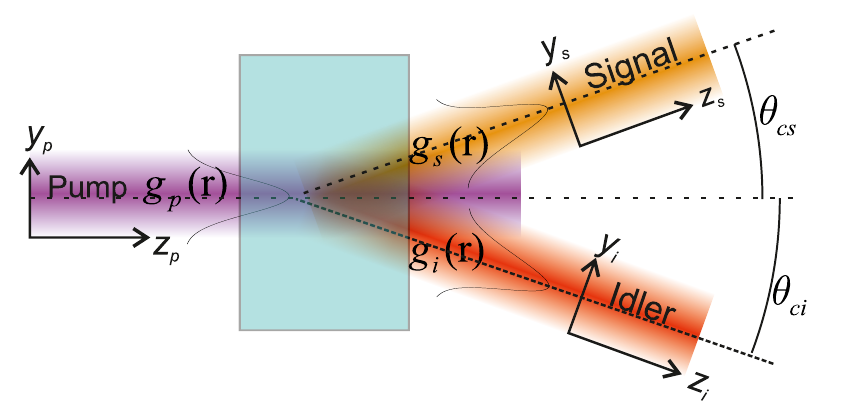}
	\caption{(Color online) The pump, signal and idler gaussian modes.}
	\label{fig:Gaussian_modes}
\end{figure}

The numerical simulation of the SPDC process was evaluated based on the theoretical approach elaborated in  \cite{Ling_Lamas-Linares_Kurtsiefer_2008}. We consider only the photon pairs generated in given gaussian signal and idler modes, defined by a registration scheme:
\begin{align}\label{eqn::electric_field}
E_{j}\left(\vect{r}\right)\sim g_{j}\left(\vect{r}\right) = \exp\left(ik_j z_{j}\right)\exp\left(\frac{-\left(x^{2} + y^{2}\right)}{W_{j}^{2}}\right),
\end{align}
where $j = i,s,p$.
The pump field $g_{p}$ transverse profile is also assumed to be gaussian (Fig.~\ref{fig:Gaussian_modes}). On the one hand the biphoton emission rate
\begin{align}\label{eqn::biphoton_generation_rate}
R_{T}\propto \left|F\left(\omega_{cs},\omega_{ci},\delta\omega,\theta_{cs},\theta_{ci},W_{s},W_{i},W_{p}\right)\right|^2,
\end{align}
depends on the mode overlap integral, and on the phasematching conditions on the other. Here $\theta_{cs}, \theta_{ci}$ are the central angles of signal and idler beams, $W_{s}, W_{i}$ and $W_{p}$ are corresponding beam waists. 
The parameter $\delta\omega$ corresponds to spectral bandwidth of detection scheme.

The diffraction grating dispersion relation yields the expression for binding angles $\theta_{cs,i}$ and frequencies $\omega_{cs,i}$:
\begin{align}\label{eqn::diff_grating_relation}
\sin\theta_{0} - \sin\gamma\theta_{cs,i} = \frac{2\pi cD}{\omega_{cs,i}},
\end{align}
where $\theta_{0}$ is the incident angle of a gaussian mode conjugated with the single-mode fiber fundamental mode (Fig.~\ref{fig:Principal_setup}), $D$ is the diffraction grating groove density and $\gamma$ is the lens system magnification. The signal and idler beam waists $W_{s,i}$ are related with the wavelengths $\lambda_{s,i}$ through
\begin{align}\label{eqn::signal_idler_waists}
W_{s,i} = \Gamma W_{f}\left(\lambda_{s,i}\right),
\end{align}
where $\Gamma$ is the lens and condensing objective optical system magnification and $W_{f}$ is the fiber fundamental mode beam waist, which is wavelength dependent:
\begin{align}\label{eqn::fiber_waist}
W_{f} = \frac{\lambda}{\pi\, \rm{N.A.}}.
\end{align}
Here $\rm{N.A.}$ is the value of the fiber numerical aperture.

Substituting expressions (\ref{eqn::phase_matching}),(\ref{eqn::diff_grating_relation})--(\ref{eqn::fiber_waist}) to (\ref{eqn::biphoton_generation_rate}) one can obtain the frequency dependence of biphoton emission rate $R_{T}\left(\omega_{cs}\right)$ governed by experimentally tuned parameters: $\gamma$, $\Gamma$ and $W_{p}$.

\section{Calculation results}

\begin{figure}
	\centering
		\includegraphics{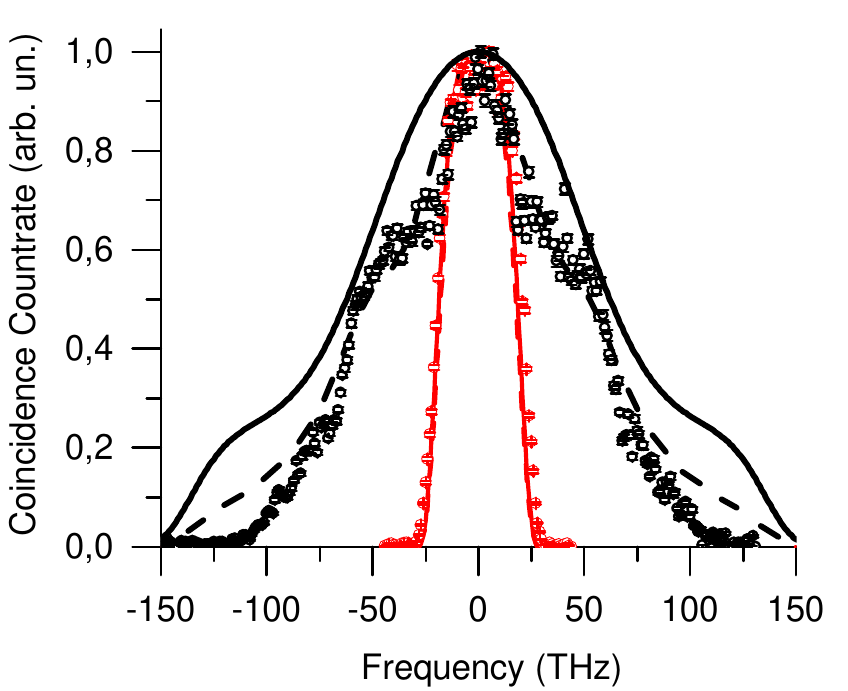}
	\caption{(Color online) Biphoton frequency spectra.
	 The value $\nu=0$ corresponds to the central frequency ${\omega_p}/{4\pi}$.
	 Solid lines correspond to ideal calculation, dashed line corresponds to exact calculation,taking into account the diffraction efficiency, circles correspond to experimental data. Black thick lines and circles refer to transformed spectrum and red thin to initial one. 
	 }
	\label{fig:Frequency_spectra}
\end{figure}

Numerical calculations were performed for the nonlinear 2 mm long BBO crystal cut for a collinear degenerate type-I phasematching, pumped by the 325~nm laser beam with a waist $W_{p} = 34$~\mcm. Optimal SPDC coupling conditions demand shrinking the pump beam waist to smaller values but, taking into account that for given nonlinear media spatial walk-off is equal to 50~\mcm, further 
pump beam waist reduction
won't 
enhance a SPDC to single-mode fiber coupling efficiency.
The diffraction grating groove density was chosen as 600/mm leading to $\gamma = 1.05$ for optimal overlap between the phasematching curve and the diffraction grating dispersion curve. According to \cite{Ling_Lamas-Linares_Kurtsiefer_2008}, to achieve the maximum collecting efficiency $\Gamma$ has to be selected in order to satisfy the condition
\begin{align}\label{eqn::beam_waist_relation}
W_{s} = W_{i} = \sqrt{2} W_{p}=48~\mu\rm{m},
\end{align}
which corresponds to the angular bandwidth $\Delta \theta=0.5^{\circ}$ in Fig.~\ref{fig:XXX}.
Considering the spectral dependence of $W_{s,i}$, $\Gamma$ was selected to satisfy (\ref{eqn::beam_waist_relation}) only for the central wavelength (650~nm). Initial and transformed calculated coincidence spectra are plotted in Fig.~\ref{fig:Frequency_spectra} (straight curve).
%
%
%
%

We used a diffraction grating blazed at 750~nm in our experiment. The first order diffraction conversion coefficient at 750~nm was equal to 70\% and drastically dropped down to 25\% at 500 nm. We introduced diffraction efficiency to our numerical model; the result is plotted by a dashed curve in Fig.~\ref{fig:Frequency_spectra}.
\begin{figure}
	\centering
		\includegraphics{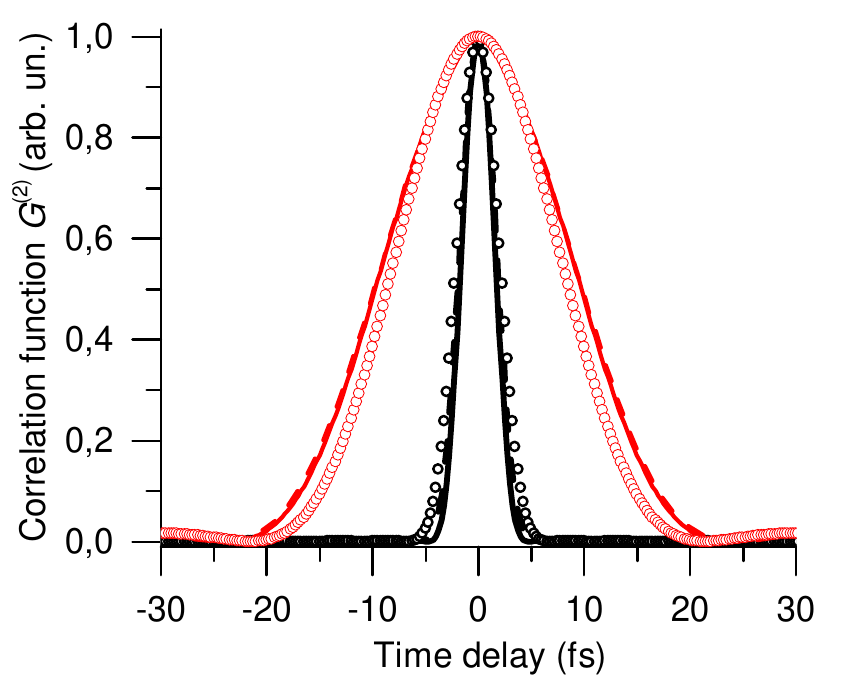}
	\caption{(Color online) Biphoton correlation functions $G^{(2)}$ \eqref{eqn::G2} calculated from the spectra plotted in Fig.~\ref{fig:Frequency_spectra}. 
		 Solid lines correspond to ideal calculation, dashed line corresponds to exact calculation,taking into account the diffraction efficiency, circles correspond to experimental data. Black thick lines and circles refer to transformed spectrum and red thin to initial one. 
	}
	\label{fig:Correlation_functions}
\end{figure}
%
%
\begin{table}
\caption{\label{tab:legend}%
The
spectral bandwidth $\Delta \nu$ \eqref{eqn::delta} and the second order correlation time $\Delta \tau$ for the plots in Fig.~\ref{fig:Frequency_spectra} and Fig.~\ref{fig:Correlation_functions}.}
\begin{ruledtabular}
\begin{tabular}{l|cc|ccc}
&\multicolumn{2}{c|}{Initial}&\multicolumn{3}{c}{Transformed}\\
&\multicolumn{1}{c}{Theory}&\multicolumn{1}{c|}{Exp.}&\multicolumn{2}{c}{Theory}&\multicolumn{1}{c}{Exp.}\\
&&&Ideal&Exact\\
\hline
$\Delta\nu$ (THz)
&36&39&144&112&99\\
$\Delta\tau$ (fs)
&20.3&18.2&3.7&4.1&4.8
\end{tabular}
\end{ruledtabular}
\end{table}
The values of the spectral bandwidths, determined as an integral characteristic:
\begin{align}\label{eqn::delta}
\Delta\nu = \int R(\nu) d\nu /R(0),
\end{align}
are presented in Table~\ref{tab:legend}. One may note that spectral dependence of diffraction efficiency leads to narrowing of the bandwidth of transformed spectrum from 144 to 112~THz.

Next we evaluated the corresponding second order correlation functions
\begin{align}\label{eqn::G2}
G^{(2)}(\tau)\propto\left|\int F(\omega) e^{-i\omega\tau} d\omega\right|^2,
\end{align}
which are presented in Fig.~\ref{fig:Correlation_functions}. The corresponding correlation times are shown in Table~\ref{tab:legend}. The correlation times have been calculated similarly to \eqref{eqn::delta}.
Ideally, 3.7~fs correlation time can be achieved.
It corresponds to 1.7 optical cycles.
Non-uniform grating spectral efficiency leads to longer 4.1~fs correlation time.

\begin{figure}
	\centering
		\includegraphics{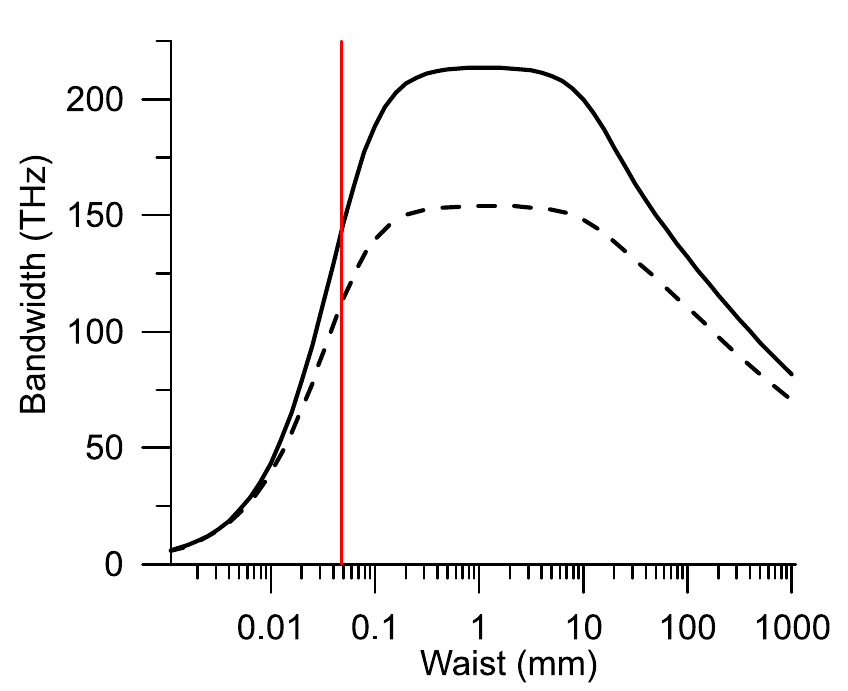}
	\caption{(Color online) The frequency bandwidth \eqref{eqn::delta} dependence on the beam waist. Straight line -- the ideal case, dashed line --  taking into account the spectral dependence of the grating efficiency. Vertical line -- the waist value $W=48$~\mcm{} used in the calculations and in the experiment.}
	\label{fig:Bandwidth_vs_Waist}
\end{figure}

We calculated the spectral bandwidth \eqref{eqn::delta} with respect to SPDC beam waists $W_{s,i}\left(650~\rm{nm}\right)$ considering the relation for the pump beam waist (\ref{eqn::beam_waist_relation}). The results are plotted in Fig.~\ref{fig:Bandwidth_vs_Waist}.
The vertical line corresponds to $W_{s,i}\left(650~\rm{nm}\right)=48$~\mcm{}, included in our numerical model. Evidently, this value appears to be non-optimal 
considering maximum spectral broadening.
Increasing the beam waist up to 500~\mcm{} leads to broadening of the spectral bandwidth up to 213~THz in an ideal case and up to 154~THz taking into account the diffraction efficiency. On the other hand, larger beam waists imply a significant drop in coincidence count rate. According to \cite{Ling_Lamas-Linares_Kurtsiefer_2008} the coincidence count rate scales as $\propto W^{-2}$.

For beam waists smaller than 100~\mcm{} our numerical model predicts a decrease in the spectral bandwidth of the coincidence count rate. This effect is associated with shrinking the interaction volume for a noncollinear generation regime in a nonlinear medium between the pump, signal and idler gaussian beams, which leads to depletion of the spectral intensity of distant spectral components. At the same time, the collection angular range increases allowing the vertical branch of the angular-frequency spectrum (Fig.~\ref{fig:XXX} c) to enhance the central spectral components and, hence, reduce the spectral width. Here we have to note that our model doesn't apply to a tight-focus scenario, when the waist length is shorter than the crystal length. Moreover, it doesn't account for broadening of the SPDC spectrum due to pump focusing \cite{Carrasco_Nasr_Sergienko_Saleh_Teich_Torres_Torner_2006, Carrasco_Sergienko_Saleh_Teich_Torres_Torner_2006, Carrasco_Torres_Torner_Sergienko_Saleh_Teich_2004b}.

Spectrum bandwidth reduction due to the beam waist decreasing is associated with angular range $\Delta\theta$ narrowing, where the phasematching curve overlaps the grating dispersion curve. Hence, the more precise overlap can be achieved for the narrower spectral range.

\section{Experiment}

\begin{figure}
	\centering
		\includegraphics{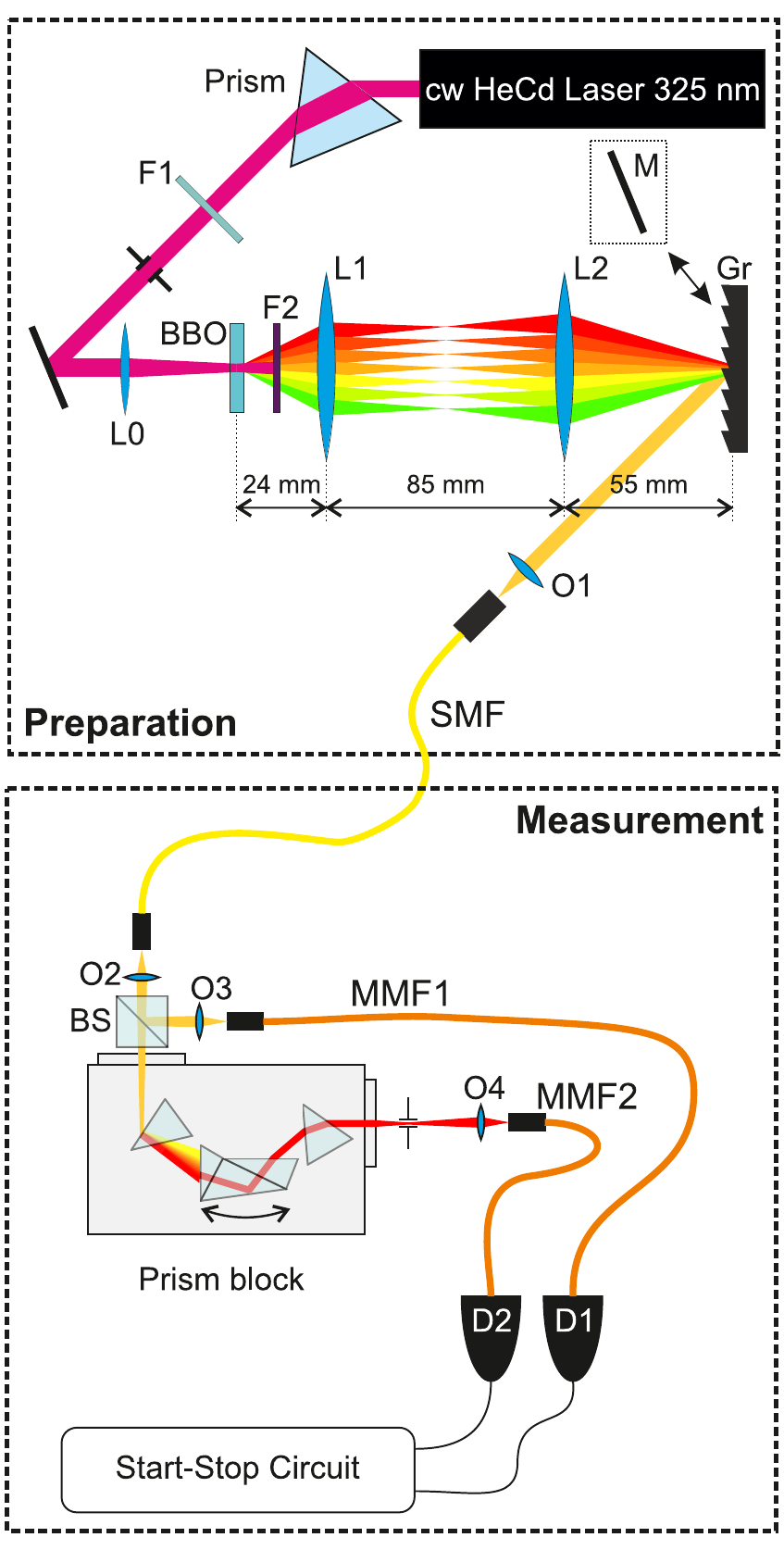}
	\caption{(Color online) The scheme of the experimental setup.}
	\label{fig:Setup}
\end{figure}

Fig.~\ref{fig:Setup} illustrates the scheme of the experimental setup.
Laser radiation at 325~nm generated by the HeCd laser was focused by a lens L0 ($f_0 = 145$~mm)  on the 2~mm long BBO crystal cut for collinear degenerate type I phasematching. A dispersive prism and filter F1 blocked parasite arc radiation. The pump beam was blocked with a UV Filter F2. Biphotons were focused through a double-lens system L1 and L2 ($f_1=f_2 = 40$~mm) on the diffraction grating, described in the previous section. The lens positions (noted in Fig.~\ref{fig:Setup} in mm) were selected to set the magnification of the system to $\gamma=1.05$.
The numerical aperture of the lens system exceeded 0.17 which corresponds to $\pm 9.5^\circ$ angular range. 
It corresponds to the spectral mismatch interval from -150 to 460~THz for the given the phasematching curve (Fig.~\ref{fig:XXX} a).
To measure an initial spectrum, the diffraction grating was replaced by a broadband mirror M.

We used a 20X objective lens O1 to couple first-order diffracted light to a single-mode fiber SMF (Thorlabs SM600). This fiber retains single-mode propagation in the 600--800~nm band. For shorter wavelengths of 500--600~nm the fiber maintains a few-mode lossless propagation, which may affect the coupling efficiency in the specified spectral range. For our scheme the effect of a non-single-mode coupling is negligible. However, for the near-infrared range starting from 800~nm the propagation losses become significant. That is why we only used a 1~m long fiber patch-cord to minimize propagation losses.

Single-mode fiber output radiation was collimated by the objective O2 and split by a 50:50 nonpolarizing broadband beam splitter BS and coupled with objectives O3 and O4 to multimode fibers MMF1 and MMF2  connected with Single-photon counting modules D1 and D2 based on Si avalanche photodiodes. One of the channels after the beam splitter included a spectrograph prism block, capable of $\delta\nu = 2\div 4 $ THz (wavelength dependent) spectral band filtering.
Sufficiently wide spectral filtering
enables both 
high coincidence count rate and 
sufficient spectral resolution.

Experimental data 
took
into account all spectral dependent parameters of the measurement part: detector quantum efficiency and wavelength dependent spectral filtering bandwidth. Thus, the key non-ideality of the preparation part (grating efficiency) was taken into account in the above theoretical model.

\section{Experimental results}

The measured transformed coincidence spectrum is plotted with  thick (black) circles in Fig.\ref{fig:Frequency_spectra}. Thin (red) circles refer to the initial spectrum measured  after placing the broadband mirror instead of the diffraction grating.
All the plots are normalized against their maximal values. The measured spectral intensities at the central frequency for initial and transformed spectra were different, which can be explained by the distinction between the grating efficiency and mirror reflection coefficient.

Agreement between the experimental data and numerical simulation (accounting for diffraction grating efficiency) is observed. The 99~THz spectral bandwidth \eqref{eqn::delta} was obtained. The signal drop on the edges of the spectral range is attributed to geometric lens aberrations. Notably, the measured initial spectrum looks slightly broader than the calculated one. It can be explained by the small broadening due to pump focusing  \cite{Carrasco_Nasr_Sergienko_Saleh_Teich_Torres_Torner_2006, Carrasco_Sergienko_Saleh_Teich_Torres_Torner_2006, Carrasco_Torres_Torner_Sergienko_Saleh_Teich_2004b}, which wasn't taken into account in the theory.

The second-order correlation functions \eqref{eqn::G2} calculated from the experimental data are also plotted with circles in  Fig.~\ref{fig:Correlation_functions}. The correlation time related to the transformed spectrum is 4.8~fs. All spectral bandwidths $\Delta\nu$ and correlation times $\delta\tau$ are presented in Table~\ref{tab:legend} in the "Exp." columns.

The impact of the single-photon impurity is quantified by the coincidence to single count rates ratio: $1-Rc/Rs$. We experimentally estimated the ratio $Rc/Rs$ (at the central frequency) to be $2.5$\%.
The low biphoton intake is attributed to significant optical losses in the measurement part of the setup and an inefficient fiber coupling scheme. This is a trade-off between the maximal broadening and maximal efficiency. Thus, the maximum possible value of $Rc/Rs$ is $45\pm 5$\% assuming any other losses are negligible. The preparation part of the setup solely introduces optical losses reducing the $Rs/Rc$ ratio down to $17\pm2$\%.



We measured coincidence count rates $R_{in} = 33$~Hz and $R_{tr} = 19$~Hz at the central frequency ($\delta\nu=3.2$~THz) for initial and transformed biphoton fields to ensure the biphoton spectral power density invariance. The ideal scenario implies $R_{tr}/R_{in}=1$. The measured ratio $R_{tr}/R_{in}=0.58$ may be attributed to a significant difference between the diffraction grating efficiency and the broadband mirror reflectance at the given frequency. Moreover, a discrepancy between the measured value of  $R_{tr}/R_{in}$ and the corresponding broadening factor $\Delta\nu_{in}/\Delta\nu_{tr}=0.39$ endorses the fact that in our experiment the spectral intensity drop isn't induced by biphoton spectral broadening.


\section{Discussion}

The demonstrated biphoton source engineering technique possesses several distinctive advantages. First, the method provides an ultra-broadband biphoton generation regime. The measured 100~THz bandwidth and 5~fs correlation time corresponds to values enabled by the homogeneous broadening techniques \cite{Shaked2014, Nasr_Giuseppe_Saleh_Sergienko_Teich_2005,ODonnell2007}. The optimization of our scheme enables bandwidth up to 144~THz, which have been demonstrated by using the inhomogeneous \cite{Katamadze_Kulik_2011} or extra-thin \cite{Katamadze_Borshchevskaya_Dyakonov_Paterova_Kulik_2013} nonlinear crystals.      



On the other hand, our method is limited by the spectral region, where phasematching and grating dispersion curves coincide, whereas other spectral broadening methods, based on utilizing inhomogeneous nonlinear media \cite{Carrasco_Torres_Torner_Sergienko_Saleh_Teich_2004b, Kalashnikov_Katamadze_Kulik_2009, Katamadze_Paterova_Yakimova_Balygin_Kulik_2011, Katamadze_Kulik_2011, Nasr_Carrasco_Saleh_Sergienko_Teich_Torres_Torner_Hum_Fejer_2008, Okano_Okamoto_Tanaka_Subashchandran_Takeuchi_2012}, are completely free of this limitation. At the same time, the biphoton field, generated in inhomogeneous nonlinear media, cannot be kept Fourier-limited, and additional compression techniques are required \cite{Brida_Chekhova_Degiovanni_Genovese_Kitaeva_Meda_Shumilkina_2009, Dayan_Peer_Friesem_Silberberg_2005, Peer2005}. The technique, demonstrated in our experiment, provides a Fourier-limited broadband biphoton source. Additional dispersive broadening of the correlation function may arise after passing through the lens system and single-mode fiber \cite{Chekhova_2002}. This effect can be avoided by substituting lenses with parabolic mirrors and realising the spatial mode filtering with a pinhole. Thus, our method (reversed in time) can also be applied for precise measurement of second-order correlation time by exploiting the sum-frequency generation scheme. The resolution of the such scheme is limited by the phasematching bandwidth in the nonlinear crystal.

However, our method exhibits significant optical losses, which leads to a considerable additional single-photon component in the state (\ref{eqn::biphoton_quantum_state}). The fact that the fabrication of efficient broadband diffraction gratings is barely feasible makes these kinds of losses inevitable. On the other hand, the losses emanate from non-optimal coupling biphotons to a single-mode fiber. The maximal coupling efficiency requires each biphoton spectral component to populate a single Gaussian mode. It can be realized by irradiating a nonlinear medium with a tightly focused pump beam. In our case the above condition is impossible to achieve for all frequency-angular components at the same time.

It should be noted that the demonstrated method enables broadband biphoton generation in a single spatial mode, whereas applications based on correlation function measurements demand photons of a pair to populate distinct spatial modes. 
This problem can be solved by a combination of two coherent single-mode biphoton sources \cite{Kim_Kulik_Shih_2000}.

Finally, the demonstrated technique significantly broadens the biphoton frequency spectrum without depleting the spectral intensity (and, moreover, with enhancing the total intensity) by redistributing energy between spatial biphoton modes.

\section{Conclusion}
We proposed the novel method of generating a spatial single-mode ultra-broadband biphoton field based on the frequency-angular spectrum transformation by means of the angular dispersion. The efficiency of this method was demonstrated both theoretically and experimentally and our calculations are in a significant agreement with our experimental results. The spectral bandwidth $\Delta\nu>100$~THz and the correlation time $\Delta \tau< 5$~fs can be reached. The key feature of our method -- spectral intensity invariance  --- was also demonstrated.

%
%

\begin{acknowledgments}
The authors are grateful to Stanislav Straupe for helpful discussions on biphoton spatial modes.

This work was supported by Russian Foundation of
Basic Research (projects 14-01-00557-a, 14-02-00749, and 14-02-31106-mol\_a) and in part by the European Union Seventh Framework Programme
under grant agreement no. 308803 (project BRISQ2) and NATO Project EAP.SFPP 984397. N.~Borshchevskaya is grateful for a support to Russian Science Foundation (project 14-12-01338).
\end{acknowledgments}

\bibliography{bibliography_paper}

\end{document}